# Designing human and generative AI collaboration: Trade-offs in productivity, satisfaction, and creative diversity

**Kartik Hosanagar[1]* and Daehwan Ahn[2]**

[1]Operations, Information and Decisions Department, The Wharton School, University of Pennsylvania, Philadelphia, PA 19104, USA
[2]Financial Planning, Housing and Consumer Economics Department, University of Georgia, Athens, GA 30602, USA
*To whom correspondence should be addressed: Email: kartikh@wharton.upenn.edu

## Abstract

**Rapid advances in generative AI have raised hopes for major productivity gains but recent research has also raised concerns about declining worker satisfaction and reduced creative diversity. While prior research has documented each of these effects individually, it remains unclear how collaboration design shapes the costs that accompany AI-enabled productivity gains. We address this question through a human-subjects experiment on a 1,000-word creative writing task, comparing three human–AI collaboration designs. Completion time was similar across the three designs but quality, satisfaction, and diversity differed across them. The Human Confirmation design, in which AI led ideation, outlining, drafting, and editing while humans primarily confirmed or rejected output, showed the clearest downsides: lower quality and satisfaction, and lower aggregate genre and semantic diversity. By contrast, designs that preserved human involvement in preproduction or throughout the writing process avoided the clearest quality and satisfaction losses without requiring longer overall completion times. Preserving active human involvement in preproduction was also associated with lower aggregate similarity in content than Human Confirmation. These findings show that collaboration design substantially reshapes trade-offs between efficiency, quality, satisfaction, and diversity, suggesting that the central challenge is not whether to adopt AI, but how to allocate creative control between humans and AI. That choice shapes not only the output of creative work but the experience of creative agency.**

### Significance Statement

AI integration is redefining the future of work, yet concerns persist about trade-offs between productivity gains and costs, including reduced work satisfaction and homogenized creative output. Studies have examined these outcomes separately, leaving unclear how collaboration design shapes these trade-offs. Comparing collaboration models experimentally, we show that how work is allocated between humans and AI can reshape trade-offs. When humans guide early creative decisions and AI supports execution, organizations can configure the trade-offs among efficiency, quality, satisfaction, and creative diversity. When AI leads and humans mainly prompt and approve, satisfaction is lower than under the other collaboration designs, and some quality dimensions are lower. Collaboration design shapes not only the output of creative work but the experience of creative agency.

## Introduction

Advances in generative AI have transformed how knowledge and creative work are performed. A growing body of evidence shows that AI tools can substantially increase productivity across domains ranging from software development (1, 2) to consulting (3), customer support (4), sales (5), and professional writing (6). In some cases, AI assistance also improves quality, particularly for lower-skilled workers who benefit most from AI assistance (6, 7). Yet these gains come with trade-offs. Recent studies raise concerns that generative AI reduces the diversity of creative outputs (7–9), risks undermining worker satisfaction (10), and may disadvantage highly skilled individuals whose expertise is displaced by automated systems (11). This tension raises a central question for organizations: how can we integrate AI to capture its productivity and quality benefits while avoiding losses in creative diversity and human satisfaction?

We argue that the answer lies in the design of human–AI collaboration. Consider human satisfaction, for example. The human–AI collaboration design can change not only the quality of the final artifact but also the way people experience the creative process. If dissatisfaction partly reflects the experience of working with AI, then collaboration design may help preserve satisfaction while still allowing AI to contribute to productivity and output quality. Existing studies on AI and human productivity often treat collaboration design as an exogenous element. Further, the collaboration design itself varies across them: sometimes AI performs all tasks while humans prompt and approve AI output (9, 12), sometimes AI acts as a copilot (1), and, in other studies, tasks are partitioned between humans and AI (5). Prior work has shown that these choices matter; certain designs, such as "ghostwriter" modes where humans only approve AI drafts, can harm experts or reduce novelty (11). But no study has systematically compared multiple collaboration designs across a comprehensive set of outcomes. We do so to investigate how collaboration can be designed to shape the trade-offs among competing objectives in creative work. One contribution of this paper is therefore to show that collaboration design affects not only the output of creative work but also the experience of creative agency.

We ground our study in the literature on creativity, which conceptualizes the process in two phases: preproduction (ideation and outlining) and production (drafting and refining/editing) (13, 14). This process-centric lens allows us to operationalize and compare three collaboration designs: Human Confirmation, in which AI leads both preproduction and production while humans only retain final approval (9, 12); Human Creativity, in which humans lead preproduction while AI handles drafting and editing (5); and Copilot, in which humans and AI collaborate continuously across both phases (1). Across all three designs, humans retain control over prompting and the authority to accept or reject AI-generated output. We interpret these treatments as positions on a continuum of human creative participation. Human Confirmation gives the human very limited creative control beyond final approval. Human Creativity front-loads human creative control in preproduction (ideation and outlining) before AI handles drafting and editing (with humans having only an approval role in that phase). Copilot preserves continuous human–AI collaboration across both preproduction and production. These designs recur across prior studies of human–AI collaboration. Table S13 maps the collaboration designs onto the preproduction/production framework.

To study these designs, we conducted an experiment with 285 participants who were asked to complete two 1,000-word writing tasks: one without AI and one with AI under a randomly assigned collaboration model. We evaluated submitted stories using human raters and surveys, examining four key outcomes: task efficiency, story quality, user satisfaction, and creative diversity. This multidimensional approach allows us to assess how different collaboration designs create trade-offs across metrics.

# Theoretical background and experimental design

Creativity is commonly defined as the production of work that is both novel and valuable (15, 16). Many frameworks exist for mapping the creative work process (17, 18). A common approach divides it into two phases: preproduction (sometimes referred to as design or generation) and production (execution or implementation) (13, 14). The preproduction phase consists of ideation and planning/structuring, while the production phase involves implementing the plan through prototyping and review/refinement (17, 19). Although both phases incorporate divergent and convergent thinking, which are central to creativity, their predominant cognitive activities differ. The preproduction phase emphasizes divergent thinking (20, 21), characterized by generating multiple ideas and exploring possibilities, which are essential for developing novelty. This phase marks the origination of innovation and aligns closely with the theoretical understanding of creativity as the process of generating novel concepts. In contrast, the production phase prioritizes organizing, refining, and enhancing the value of pre-existing ideas (22). While this phase plays a critical role in ensuring the utility and coherence of creative outputs, it contributes less to the generation of original ideas. In creative writing, the process can be divided into four key steps: ideation and outlining, which fall under the preproduction phase, and drafting and editing, which are part of the production phase (21, 23–25).

Building on this framework, we conducted an IRB-approved classroom experiment to examine the effects of different AI collaboration models during preproduction and production phases of creative writing. 285 students participated in a 2-day study, writing 1,000-word stories without AI (day 1) and with AI assistance (day 2). We study a single-session, single-artifact writing task because its preproduction and production stages could be separated and timed within one session, making it easier to manipulate where in the process humans retained creative control.

The study protocol was approved by the University of Georgia Human Research Protection Program (IRB ID: PROJECT00008677). Before the study, students received a consent letter describing the study procedures, the voluntary nature of participation, their right to decline or withdraw at any time without penalty, and the anonymization of all research records; students indicated their decision to participate to the instructor, and informed consent was obtained from all participants. All participants received the same bonus credit regardless of condition, and students who chose not to participate could complete an alternative task for the same credit.

On day 1, participants completed an initial survey assessing participant heterogeneity with regard to their English proficiency, creative writing skills and experience, and a Divergent Association Task (DAT) test to measure their verbal creativity (26). They then wrote a 1,000-word story without any AI assistance during a session that lasted 105 minutes, following a seven-sequence structure commonly used in fiction writing [Supporting Information (SI) Section B.2.4]. Upon completion, they filled out an exit survey that captured the time spent on each writing stage, their writing experience (the flexibility and effectiveness of the process), and their satisfaction with and willingness to reuse the process (SI Section B.3.3).

On day 2, participants wrote a new 1,000-word story during a 105-min session using ChatGPT 3.5, following their assigned collaboration model. Participants were randomly assigned to one of three AI collaboration models. In Human Confirmation, the AI handles all four tasks (ideation, outlining, drafting, and editing) and humans confirm or reject the output. In Human Creativity,

**Table 1.** Descriptive statistics.

| Category | Variables | *N* | Mean | SD |
|---|---|---|---|---|
| Personal Characteristics (Initial Survey) | DAT Score | 280 | 82.28 | 7.13 |
| | Creative Writing Experience | 280 | 1.62 | 0.71 |
| | Creative Writing Ability | 280 | 1.90 | 0.67 |
| | English Language Proficiency | 280 | 2.77 | 0.43 |
| Quality Metrics (Rated by 4 Evaluators per Story) | Overall Quality | 2,196 | 4.91 | 1.35 |
| | Originality | 2,196 | 4.97 | 1.38 |
| | Interestingness | 2,196 | 4.87 | 1.46 |
| | Writing Quality | 2,196 | 5.21 | 1.37 |
| | Coherence | 2,196 | 5.20 | 1.36 |
| Satisfaction with the Writing Process (Exit Surveys on day 1 and 2) | Process Satisfaction | 508 | 4.90 | 1.53 |
| | Flexibility | 508 | 4.95 | 1.47 |
| | Process Effectiveness | 508 | 4.73 | 1.57 |
| | Intent to Reuse Process | 508 | 0.72 | 0.45 |
| Satisfaction with the AI Support (Exit Survey on day 2) | Satisfaction with AI | 248 | 5.47 | 1.42 |
| | AI Effectiveness | 248 | 5.23 | 1.65 |
| | AI's Contribution | 248 | 0.76 | 0.17 |

This table presents descriptive statistics for our dataset. *N* denotes participants for Personal Characteristics and Satisfaction with AI Support, evaluator-story ratings for Quality Metrics, and participant-day responses for Satisfaction with the Writing Process.

humans perform preproduction (ideation and outlining) without AI assistance, while the AI executes the production phase (drafting and editing) with humans confirming or rejecting AI output as in the Human Confirmation design. In Copilot, humans and AI collaborate throughout the entire process from ideation all the way to editing. Detailed instructions for each group are provided in SI Section B.2 and Table S13 summarizes how each design allocates preproduction and production tasks between humans and AI. Each design is implemented in a deliberately strict form. Participants in Human Confirmation, for example, were instructed not to offer additional ideas at any stage of the process. The strictness is intended to isolate where human creative control sits within the workflow. While other potential designs could theoretically be explored, we selected these three because they capture theoretically distinct allocations of human creative control that appear in prior research (5, 9, 12, 25). Our objective is not to reproduce the full complexity of naturally occurring workflows, but to isolate the consequences of shifting creative control across stages of the process. After receiving training in prompt engineering techniques to address AI-skill disparities, participants completed the task and an exit survey identical to day 1, with additional questions about AI effectiveness, collaboration satisfaction, and AI's contribution (SI Section B.3.4). Participants also submitted a 2-page reflection on their collaboration experience along with their full ChatGPT interaction history (SI Sections A.1 and C.1).

The study focused on four main outcomes: completion time, story quality, user satisfaction, and content diversity. Story quality was evaluated by two internal (fellow participants) and two external raters (recruited from the University) using a 1–7 Likert scale for overall quality, originality, interestingness, writing quality, and coherence. Completion time and user satisfaction were measured through exit surveys (SI Sections B.3.3 and B.3.4). Content diversity was analyzed by comparing story similarities and genres across AI collaboration models. Additionally, we conducted an analysis of reflective feedback to gain a deeper understanding of participants' writing experiences.

A total of 549 stories were produced, each receiving four evaluations, totaling 2,196 scores (SI Sections A.1 and B.4). To understand the impact of human–AI collaboration design, we employ a mixed between- and within-subject design. For outcomes measured on both days, our primary specification is a difference-in-differences model with participant fixed effects, which combines the benefits of random assignment with controls for time-invariant individual heterogeneity. We supplement these analyses with descriptive within-person comparisons from day 1 to day 2. Because all participants completed the non-AI task before the AI-assisted task, these within-person comparisons may also reflect learning, fatigue, or novelty effects and are interpreted more cautiously. We conducted a randomization check using fourteen baseline variables available before treatment on day 2: four participant characteristics (Creative Writing Experience, Creative Writing Ability, English Language Proficiency, and DAT Score) and 10 day 1 baseline outcomes (five quality ratings, four process-satisfaction measures, and total completion time). No differences survive Bonferroni correction. The results of the analysis are reported in Tables S0b–S0c, and participant and observation flow is reported in Fig. S0 and Table S0a. Descriptive statistics about the sample are available in Table 1. The Supporting Information includes a full description of our experiment design, relevant instruction materials and survey questionnaires, definitions of variables, and additional texts, figures, and tables validating and extending our main results.

## Results

Our analysis centers on randomized comparisons across the three day 2 collaboration designs, supplemented by within-person changes from day 1 to day 2 as descriptive context.

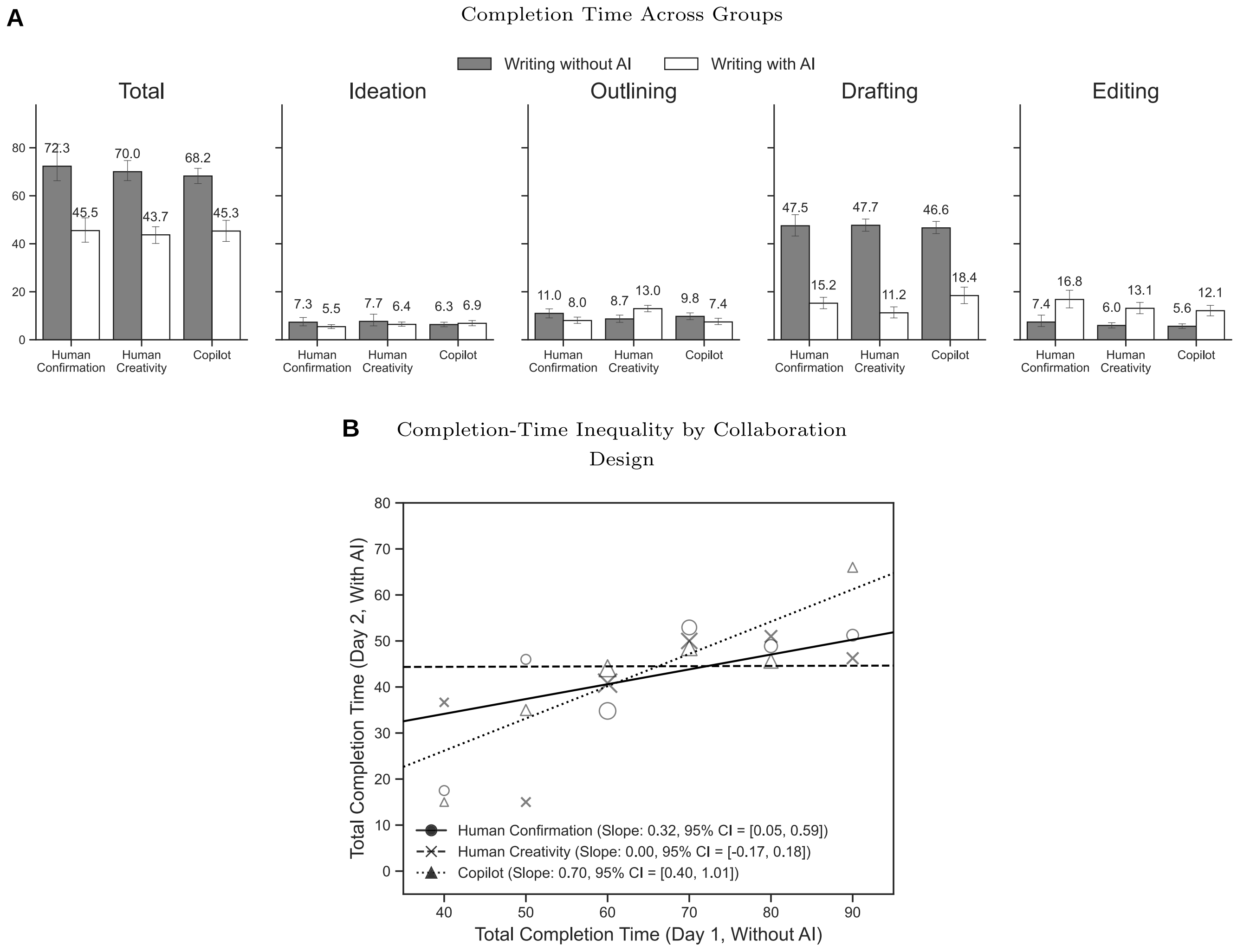

**Figure 1** Completion time by collaboration design and day. A) Graph of mean self-reported completion time (total and across writing stages) and 95% CI for collaboration groups, without AI (day 1) and with AI (day 2). B) Relationship between day 1 and day 2 completion times. The line is the regression line based on raw data for each collaboration group. The slope indicates the relationship: a slope of 1 indicates that the gaps between participants' day 1 times stayed the same on day 2, while a slope of 0 means day 2 times were unrelated to day 1 times within that design. Also plotted are observations binned by day 1 completion times, with larger markers representing bins with more observations.

## Completion time

As illustrated in Fig. 1A, total completion times were similar across the three day 2 designs, but time allocation across writing stages differed by model. The Copilot group showed smaller day 1-to-day 2 reductions in drafting time but required less editing time, while the Human Creativity group invested more time in outlining but showed a larger reduction in drafting time. Despite these variations in time allocation patterns, difference-in-differences models with participant fixed effects revealed no statistically significant differences in total completion time across groups (SI Section C.2 and Table S3). Total completion time was 36.2% lower on day 2 than on day 1, with the largest change during the drafting stage (68.6% lower). Because day 1 always preceded day 2, these within-person changes cannot separate AI assistance from prior-task learning or other order effects. Notably, editing time increased across all treatment groups, potentially reflecting either reduced involvement during drafting, which necessitated additional refinement to align output with writer preferences, or strategic reallocation of time saved in earlier stages toward more thorough editing.

We next analyzed whether participants with longer completion times without AI continued to need relatively more time when using AI assistance. The relationship between day 1 and day 2 completion times varied significantly across collaboration models. Figure 1B shows the results of a regression of day 2 completion time against day 1 completion time with controls for DAT score, writing experience, writing ability, and language proficiency. In the Copilot model, while the productivity gap modestly narrowed, the relative differences in completion times among participants remained significant ($b = 0.70$, $P < 0.001$). The Human Creativity model, however, eliminated these productivity differences ($b = 0$, $P = 0.956$), with participants achieving similar completion times on day 2 regardless of their day 1 performance. This equalization may be because participants' main role in this model is to drive the early creative tasks, which anyway constitute a relatively small portion of the total completion time. The Human Confirmation model demonstrated an intermediate effect ($b = 0.32$, $P < 0.05$), showing a weaker but still significant relationship between day 1 and day 2 completion times.

## Writing quality

As illustrated in Fig. 2A, on day 1 (without AI assistance), story quality showed no statistically significant differences across treatment groups (SI Table S0b). However, a participant-fixed-effects difference-in-differences model showed that the day 1-to-day 2 change in overall quality was lower under Human Confirmation than under Copilot ($b = -0.43$, $P = 0.007$). The corresponding contrast with Human Creativity was smaller ($b = -0.271$, $P = 0.079$), not statistically significant at the 0.05 level. The day 1-to-day 2 differences for Human Creativity and Copilot were not statistically distinguishable from one another. The full set of contrasts is reported in Table S14. These findings suggest that active human participation in the preproduction tasks of writing, namely ideation and outlining, is important for maintaining quality when using AI assistance. The lower overall quality under Human Confirmation stems primarily from reduced interestingness and coherence (Fig. 2C). Within-person paired tests show the same pattern in absolute terms: under Human Confirmation, interestingness fell below the no-AI baseline (4.96 to 4.72; paired-*t* $P = 0.046$, $d = -0.22$), and overall quality and coherence also declined but were not statistically significant, whereas Human Creativity and Copilot showed no such declines (SI Table S15).

Quality gaps associated with baseline writing performance also differed across the three designs. A regression of day 2 Overall Quality on day 1 Overall Quality (with controls for DAT score, writing experience, writing ability, and language proficiency) showed that writers with stronger day 1 (no-AI) performance were less effective under Human Confirmation than under Human Creativity or Copilot, where human preproduction was preserved (Figure 2B). DAT scores provide an alternative measure of participant creativity. The results are qualitatively similar with this measure. As shown in Fig. 2D, participants with higher DAT scores initially produced higher quality writing on day 1, without AI assistance ($b = 0.014$, $P < 0.05$). On day 2, the association between DAT score and quality was not statistically distinguishable from zero ($b = -0.001$, $P = 0.933$). Figure 2E shows the day 2 quality pattern across DAT groups. Writers in the Average and High DAT groups produced higher-quality stories under Human Creativity and Copilot than under Human Confirmation. Full estimates in SI Section A.3.

## User satisfaction

Participant satisfaction exhibited significant variation across AI collaboration models (Fig. 3A). The Human Creativity and Copilot groups reported higher levels of satisfaction, perceived flexibility, and effectiveness on day 2 compared to the Human Confirmation group. From day 1 to day 2, perceived flexibility showed the clearest within-person decline. Under Human Confirmation it fell from 5.10 to 4.31 (paired-*t* $P = 0.002$, $d = -0.39$), with no comparable decline under Human Creativity or Copilot. The cross-group difference-in-differences contrasts with participant fixed effects are negative but not statistically distinguishable from zero (Table S17). Overall process satisfaction did not improve under Human Confirmation, whereas Human Creativity and Copilot improved on several other process items (Table S16). Notably, while the Human Confirmation group acknowledged greater AI contribution to their final output, they reported lower ratings for both AI effectiveness and satisfaction with AI assistance (Fig. 3B). These findings suggest that preserving meaningful human creative control is associated with greater satisfaction with AI-assisted writing. In our treatments, this control was concentrated in ideation and outlining under Human Creativity and distributed throughout the workflow under Copilot.

## Content diversity

GPT-4o was employed to classify all stories into one or more of nine popular literary fiction genres (27). Among the three day 2 designs, the number of genres incorporated in each story rose with the amount of creative work assigned to AI, indicating more multifaceted genre composition. The Human Confirmation group showed the highest average number of genres per story (mean = 4.07, 95% CI = [3.90, 4.23]), followed by Copilot (mean = 3.76, 95% CI = [3.60, 3.92]), Human Creativity (mean = 3.55, 95% CI = [3.40, 3.69]), and writing without AI (mean = 3.45, 95% CI = [3.36, 3.53]).

An analysis of story genres revealed notable differences across conditions (Fig. 4A). Young Adult fiction was most common in the day 1 stories and less common in the conditions where AI took on more creative tasks, particularly in the Human Confirmation group. Meanwhile, genres requiring more imaginative elements such as Fantasy, Mystery/Crime, and Sci-Fi were more common in the AI-assisted conditions. Across the randomized day 2 designs, stories relied less on personal experience-based narratives and more on these popular genres where more creative work was assigned to AI, likely reflecting their prevalence in AI training data. As a partial sensitivity check, an independent reclassification with Qwen2.5-72B-Instruct (a non-OpenAI open-weight model) supports the broad group-level genre pattern (Table S18).

To assess the variance of genre composition within any group, we computed the genre similarity between stories. Genre similarity was calculated using cosine similarity between each story's genre vector—a vector comprising nine binary values where a genre classification is indicated by 1 and others by 0—and the average genre vector of other stories within the same group. Story genres are primarily determined during ideation and outlining in the preproduction phase. Figure 4B demonstrates that when humans were excluded from preproduction, genre similarity increased within groups. The Human Confirmation group demonstrated significantly higher genre similarity (mean = 0.592, 95% CI = [0.585, 0.598]) compared to writing without AI (mean = 0.532, 95% CI = [0.530, 0.534]). In contrast, other conditions which continued to involve human participation in ideation and outlining showed no statistically significant increases in genre similarity compared to writing without AI (Human Creativity: mean = 0.533, 95% CI = [0.526, 0.539]; Copilot: mean = 0.527, 95% CI = [0.521, 0.534]).

Figure 4C shows an analysis of story similarities within each treatment group. These were calculated using cosine similarity between each story's embedding and the average embedding of other stories in its group, based on OpenAI's "text-embedding-3-large" model (28). The Human Confirmation group, where AI had the greatest role, showed the highest similarity (mean = 0.779, 95% CI = [0.767, 0.790]), followed by Copilot (mean = 0.743, 95% CI = [0.731, 0.755]), Human Creativity (mean = 0.710, 95% CI = [0.700, 0.720]), and writing without AI (mean = 0.647, 95% CI = [0.642, 0.653]). These differences were statistically significant. All AI-assisted conditions showed higher similarity than the day 1 stories, and among the randomized day 2 designs similarity was highest where AI led ideation. This result aligns with previous research (7), which showed that AI-generated stories tend to be more alike than those created by humans. However, our analysis additionally highlights that managers and creators can influence this outcome by adjusting collaboration design. Specifically, the two designs that retained human involvement in preproduction showed lower semantic similarity and higher day 2 story quality than Human Confirmation, while their day 1-to-day 2 changes in completion time did not differ from Human Confirmation (Tables S3 and S7). Among AI-assisted groups, the Human Creativity model produced the most diverse stories, demonstrating the value of preserving human input in the preproduction process. Figure 4A–C together suggests that, although Human Confirmation stories draw on a wider range of genres individually (Fig. 4A), aggregate diversity is lower (Fig. 4B and C) because different writers are nudged in similar directions by the AI.

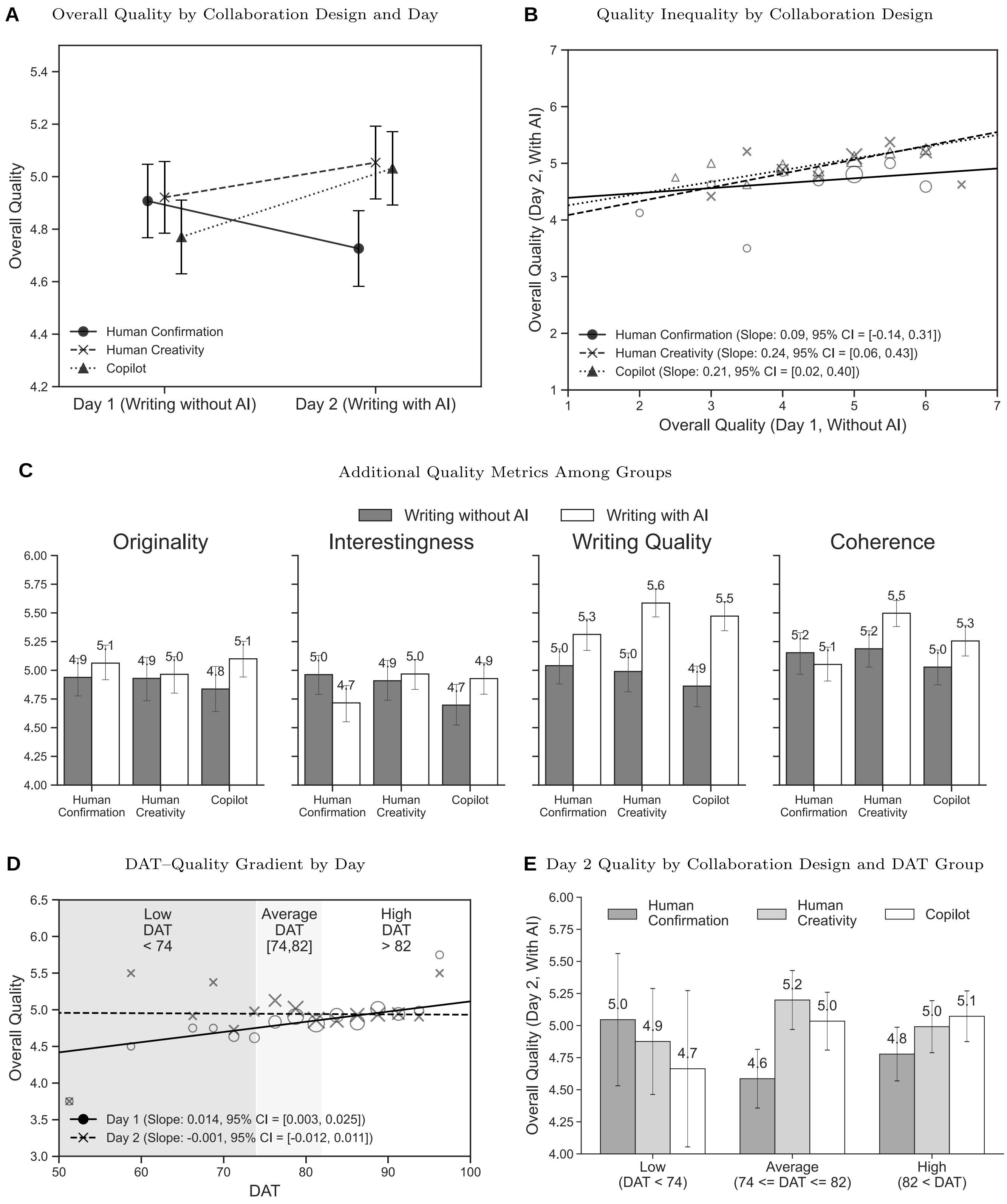

**Figure 2** Overall story quality by collaboration design and day. A) Means and 95% CIs for overall story quality by collaboration group, without AI (day 1) and with AI (day 2). B) Relationship between overall quality on day 1 and on day 2. The line is the regression line based on raw data for each collaboration group. A slope of 1 indicates that the gaps between participants' day 1 quality stayed the same on day 2, while a slope of 0 indicates day 2 quality was unrelated to day 1 quality within that design. Plotted observations are binned by day 1 overall quality. C) Mean values of additional quality metrics and 95% CIs by collaboration group, without AI (day 1) and with AI (day 2). D) Relationship between DAT score and overall story quality for each day. Observations are binned by DAT score, with the line representing a regression line based on raw, unbinned data. E) Mean values and 95% CIs of overall story quality on day 2 (with AI) across collaboration groups, categorized by low, average, and high DAT scores. With a typical range for DAT scores in the general population being 74 to 82 (26), we defined the "average DAT" group as scores within this range, and the "low" and "high" groups as below 74 and above 82, respectively.

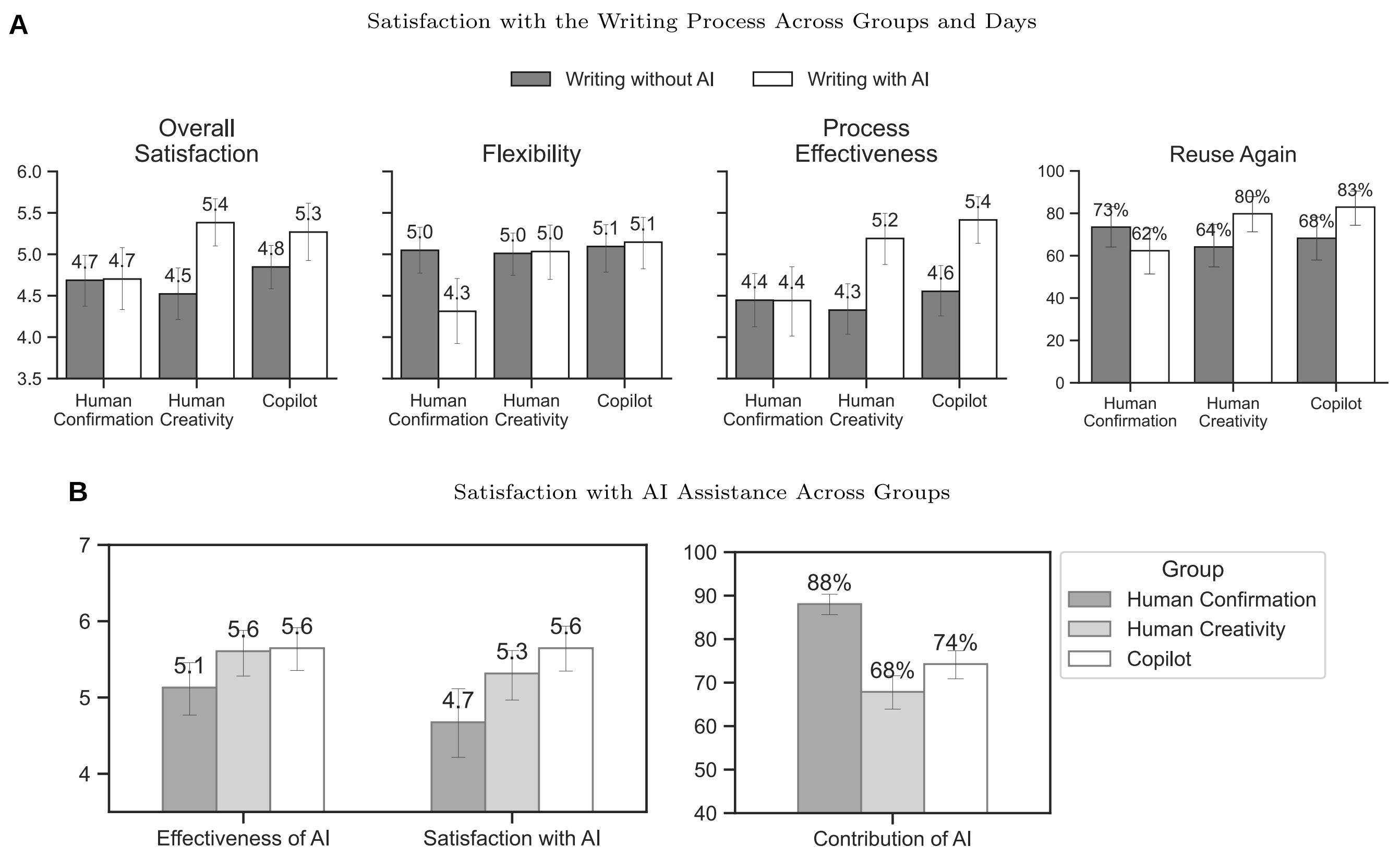

**Figure 3** Satisfaction with the writing process and with AI support. A) Participants rated their overall satisfaction, flexibility, and process effectiveness on 7-point Likert scales. They responded to three questions: "How would you rate your overall satisfaction with the writing process?," "Did the writing process allow for flexibility to achieve what you wanted?," and "Do you feel confident that you will be able to achieve your writing goals with this process?" Additionally, participants indicated whether they would use the same process again (binary Yes/No) by answering: "Did the writing process help to achieve what you wanted? Would you use the same writing process again in the future?" B) Participants evaluated aspects of AI support. They rated the effectiveness of AI assistance by answering: "How satisfied are you with the AI's help in your writing, including the quality of its suggestions and how much it boosted your productivity?" Their satisfaction with AI support was assessed through the question: "How satisfied were you with the AI-human collaboration experience you were asked to follow?" Lastly, participants quantified AI's contribution by answering: "Please specify ChatGPT's contribution to generating the story, using a scale from 0 to 100% (where 0% indicates you contributed 100%, 100% indicates ChatGPT contributed 100%, and 50% indicates an equal contribution from both)."

## Reflective feedback

In their post-study reflections, participants described what worked well and what challenges they faced when writing with AI support. Following Ludan et al. (29), we extracted the main positive and negative topics from these reflections. Each participant's reflection was then classified into these topics using Claude Sonnet 4.6. Reflections could be assigned to none, one, or multiple topics based on their content. Our analysis identified three key benefits of LLMs in creative writing: Efficiency, with AI improving structure, reducing writing time, and aiding planning; Creativity and Idea Generation, as AI helped with diverse storylines; and Language and Structural Support, enhancing grammar, coherence, and fact-checking. However, three main challenges also emerged: Authenticity and Emotional Depth, where AI lacked the human touch and emotional richness; Technical and Logical Consistency, with AI struggling to follow complex instructions and maintain a consistent voice and storyline; and Originality and Nuance, where AI produced predictable content and had difficulty with novel ideas and subtle humor. A full list of positive and negative topics mentioned in the reflective feedback (and their frequency) is provided in SI Section C.1 and Tables S1 and S2.

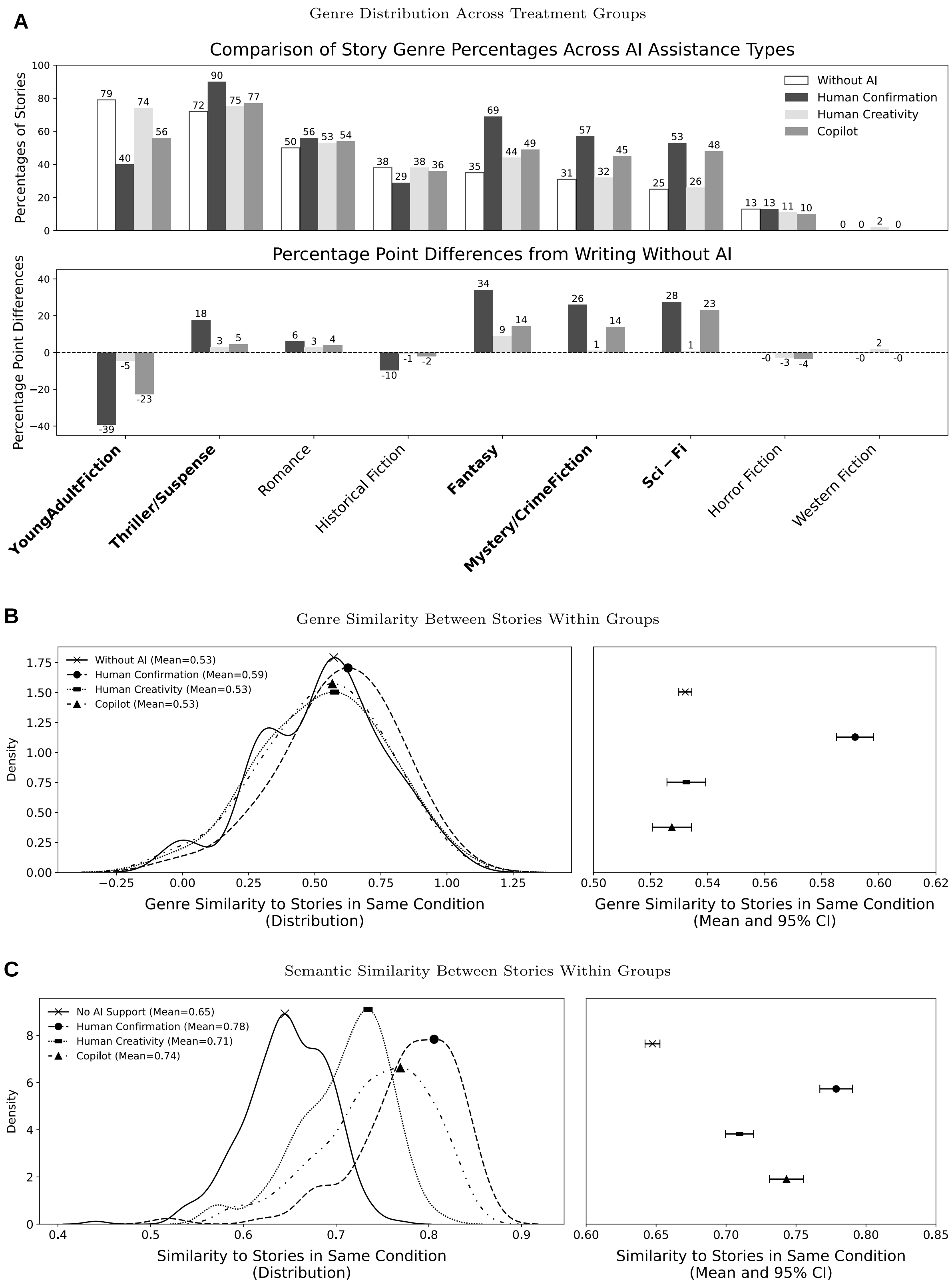

**Figure 4** Content-diversity measures across conditions. (A) Genre distributions across AI assistance types are shown as percentages (above), with differences from non-AI writing displayed as percentage point changes (below). Since stories can belong to more than one genre, the cumulative percentages exceed 100%. (B) Kernel density plots (left) and means with 95% CIs (right) illustrate genre similarity within conditions. (C) Kernel density plots (left) and means with 95% CIs (right) illustrate story (semantic) similarity within conditions. High similarities indicate that the stories share close semantic features, such as narrative structure, themes, and linguistic style.

Our analysis also revealed distinct experiential patterns across collaboration models (SI Section C.1). In the Human Confirmation group, where participants were limited to validating AI-generated outputs, the limited creative involvement was a source of dissatisfaction. Participants reported feeling "super removed from the content of the story," resulting in disengagement, diminished ownership, and final products they were "far from proud of." They struggled to guide AI narratives, noting outputs often "diverged" from their vision. Additionally, participants noted limitations in the AI's capacity for nuanced storytelling, resulting in outputs characterized as "generic," "flat," and lacking depth.

In contrast, the Human Creativity group, where participants controlled ideation and outlining, exhibited greater satisfaction and connection to their work. One participant noted that shaping the story from the start kept it "aligned with my vision," though challenges persisted in AI drafting, as "it just didn't have my voice." Still, this model fostered stronger ownership and better story quality. Similarly, the Copilot group, with continuous human–AI collaboration, reported high satisfaction and quality. Participants valued the "true partnership," leveraging AI iteratively—e.g. "suggesting tweaks for each paragraph"—to polish narratives while retaining creative autonomy. Participants emphasized process flexibility and creative autonomy.

To better understand the mechanisms underlying these differences, we conducted a structured coding analysis of the 270 post-study reflection essays (also reported in SI Section C.1). Using a predefined coding rubric, an independent large language model coded both process-related reflections (e.g. who generated the initial idea, who created the outline, and who performed drafting and editing) and experience-related themes (e.g. pride in the final story, partnership with the AI, loss of authorship, suppression of personal voice, and coherence problems). The analysis revealed that the three collaboration designs differed primarily in who controlled ideation and outlining, rather than in drafting behavior. Among participants in Human Confirmation, 80.2% explicitly indicated in their reflection that AI generated the initial idea and 54.9% indicated that it generated the outline, whereas participants in Human Creativity generated the initial idea and outline themselves in 77.7 and 77.7% of essays, respectively. Copilot occupied an intermediate position. In terms of their use of AI for drafting the stories and their use of iterative prompting, there were no notable differences. These patterns are consistent with treatment assignment. A stratified random 20% subsample of the essays ($n = 54$) was independently coded by a human coder using the same rubric. SI Table S22 reports the agreement between the human and automated codes.

The reflection coding also helps explain why Human Confirmation performed worse while Human Creativity and Copilot produced similar outcomes on many measures. Among the experience codes, the clearest differences were not in explicit complaints about personal voice or loss of authorship. They were in expressed pride and partnership with AI. Participants in Human Confirmation expressed less pride in the final story than participants in Human Creativity and Copilot combined ($P = 0.0052$, nominal) and a weaker sense of partnership with AI ($P = 0.0025$, surviving a Bonferroni correction across the 10 experience themes; Table S21). Human Creativity and Copilot did not differ on these dimensions ($P = 0.28$ for pride; $P = 0.11$ for partnership). Taken together, these findings suggest that preserving meaningful human creative control may matter more than whether that control occurs early in the process or throughout the workflow.

These findings underscore the value of AI as a supportive tool enhancing, not replacing, human creative ideation. While the specific mechanisms differ—with Human Creativity emphasizing initial creative control and Copilot enabling continuous creative partnership—both models position AI as an augmentative force that amplifies human creative capabilities rather than taking over those tasks. This combination of human creative input and AI support not only yields higher quality outputs but also maintains the engagement and satisfaction essential for sustained creative work.

# Discussion

Our study shows that how tasks are divided between humans and AI changes the trade-offs among efficiency, quality, satisfaction, and diversity. The Human Confirmation results are best read as a cautionary case about what can happen when a workflow substantially reduces the human role to prompting and approval. Whether similar costs arise in less constrained and more iterative real-world workflows remains an empirical question. More generally, this insight reframes AI integration from a question of adoption ("should we use AI?") to one of design ("how should we use it?").

By systematically comparing three collaboration models across multiple objectives, we find that each produces a distinct configuration of trade-offs. Organizations rarely optimize a single outcome in isolation. Instead, they simultaneously care about productivity, quality, employee satisfaction, learning, and creative diversity. Because these objectives are often in tension, identifying the trade-offs can enable better collaboration design. Human Confirmation, where AI drives ideation and production while humans review and accept outputs, achieves efficiency at the expense of the very dimensions that make creative work meaningful: interestingness, ownership, and variety. The same design issue can arise beyond creative writing, including software development tools that generate code for developers to select, approve, or revise. Human Creativity and Copilot both preserve meaningful human creative control. Human Creativity concentrates it in ideation and outlining, while Copilot distributes it throughout the process. Both avoid the clearest of these costs while retaining the productivity gains. They differ mainly in where the remaining trade-offs fall. Human Creativity preserves somewhat greater aggregate diversity, and both show higher writer satisfaction than Human Confirmation. The dissatisfaction articulated by participants in the Human Confirmation group evokes the IKEA effect (30), wherein individuals ascribe greater value to artifacts they have actively shaped. Our evidence also suggests that preserving meaningful creative control and effort matters; however, the stage at which that control or effort is most consequential is likely to depend on the nature of the task.

Although the IKEA effect emphasizes effort, our results are consistent with ownership and satisfaction depending on whether writers meaningfully contributed toward the creative output. This also aligns with research on algorithm aversion showing that control over algorithmic outputs significantly increases user acceptance and satisfaction (31). Importantly, whereas quality and efficiency outcomes are likely to shift with advances in AI capability, the satisfaction differences documented here likely reflect a more fundamental aspect of human psychology, namely the desire to exercise agency, judgment, and creative ownership. This may make them more durable across future generations of AI systems. The homogenization documented in the diversity results is consistent with an anchoring effect (11, 32), whereby writers who may have previously approached the creative task very differently now anchor their approach around AI's standardized suggestions, and with isomorphism (33), wherein the widespread adoption of identical AI tools fosters uniformity across organizational outputs—a concern echoed in Whittaker et al.'s (34) critique of algorithmic homogenization.

Our results highlight that collaboration design can mitigate the homogenization observed in AI-assisted writing. Where AI led ideation, individual stories incorporated the most genres, yet collective outputs converged most strongly around similar patterns and genres. Compared with Human Confirmation, Human Creativity and Copilot had lower within-condition genre and semantic similarity.

Taken together, these results reveal that collaboration design affects the trade-offs associated with AI integration. This extends recent work on generative AI in creative problem-solving (35) by showing how specific design choices shape trade-off configurations. Designs that retain human agency in early creative phases harness AI's strengths without displacing human agency, satisfaction, or the diversity of collective work.

This perspective extends existing theories of socio-technical systems (36), which argue that the success of technology implementation depends not just on the tools themselves but also on the people and processes surrounding them. These theories further hold that technology both enables and constrains human action depending on how it is embedded in social practices. Similarly, our study emphasizes design granularity, i.e. not merely choosing between "human-in-the-loop" and "full automation," but carefully designing the human–AI collaboration. In this task, the preproduction phase appears especially consequential for aggregate diversity, because it determines many initial creative choices. Satisfaction, by contrast, appears to depend more broadly on preserving meaningful human creative participation in the workflow (during preproduction in Human Creativity and throughout the process in Copilot). Meanwhile, the production phase, characterized by time-consuming implementation and refinement, benefits considerably from AI support.

Our findings also underscore the importance of adapting collaboration design to differences in baseline performance. Across all three designs, writers with lower baseline performance showed the largest gains relative to their own day 1 stories, whereas writers with stronger day 1 quality or higher DAT scores produced their best day 2 stories under the designs that let them lead the early creative tasks. Similar patterns emerge in analytical tasks, where LLMs match expert performance on structured extraction but human judgment remains essential for synthesis and interpretation (37). These patterns suggest that future work should examine whether collaboration design can be tailored by allocating greater creative control to the stronger baseline performers while still offering AI-driven support to the rest. These findings on how collaboration design relates to measured baseline performance and creativity are consistent with the task-technology fit framework (38), which asserts that technological interventions are most effective when they address specific deficiencies or unmet needs within a user's repertoire. These two findings are also reminiscent of the expertise reversal effect (39), which indicates that technological assistance that benefits novices can be ineffective, or at times detrimental, for experts. Faraj et al.'s (40) call for flexible, contingency-based frameworks finds theoretical support here, as our results highlight the value of tailoring collaboration design to the measured baseline performance among writers. For the academic community, our findings suggest the need for an agenda for developing more granular theories of AI integration that reflect differences in baseline performance and measured creativity; for managers, it suggests the potential value of deploying AI flexibly, giving writers with weaker baseline performance more support while preserving sufficient flexibility and autonomy for stronger baseline performers to maximize their creative potential.

# Limitations

We note several important limitations of our study. First, our findings focus on creative writing with AI, and caution should be exercised when generalizing to other domains such as AI use in innovation, software development or research. Second, while we used ChatGPT 3.5 (the leading language model at the time), newer models may yield different results in terms of time savings, output quality, writer satisfaction, and content diversity. Third, the study did not counterbalance the order of the two writing tasks, because all participants wrote without AI on day 1 and with AI on day 2. Within-subject changes from day 1 to day 2 may therefore also reflect practice, fatigue, or novelty, and we rely on the randomized day 2 comparisons and the difference-in-differences estimates as the most credible evidence on differences among collaboration designs. Fourth, our study was limited to 1,000-word creative writing tasks due to experimental constraints; more complex projects like writing a novel may require longer timeframes or different collaboration models. Finally, while the Human Confirmation design showed limitations in our creative setting, it may be better suited for routine, repetitive tasks that neither provide personal satisfaction nor require nuanced understanding beyond AI's capabilities.

Despite these limitations, we expect our main conclusion that collaboration design can help organizations configure the trade-offs among the competing goals of productivity, quality, satisfaction, and content diversity to extend to other creative work with distinct preproduction and production phases. For instance, in new product development, teams first engage in a design phase where they ideate and design new products, followed by a production phase involving prototyping, refining, and manufacturing. Similarly, in advertising, there is a preproduction phase focused on conceptualizing campaigns, crafting messages, and designing visuals, which is then followed by a production phase where advertisements are produced and disseminated. Separating preproduction from production as we do in our study is most useful when the main creative directions are shaped early but may be less relevant when later work substantially shapes the creative direction. Direct testing is nonetheless needed before extending these empirical claims beyond a single-session 1,000-word creative writing task. Finally, while newer LLM models may offer greater productivity gains, we focus on the relative impacts of collaboration designs, which are less likely to be affected by ongoing model improvements. Future work can extend our analysis to consider a wide range of creative tasks with a comparison of different vintages of AI models.

## Implications

The literature on work design has extensively explored collaboration within human teams. However, the design of Generative AI and human collaboration remains underexplored due to its nascency. Given the anticipated scale and scope of Generative AI's impact across various industries and jobs, studying the design of AI-human collaboration is critical and timely.

For workflows in which preproduction and production can be separated, our policy recommendations for managers emphasize that while multiple collaboration designs can yield time savings, passive human participation in early creative tasks can negatively affect both worker satisfaction and work quality. This insight underscores the need for designs that judiciously allocate tasks according to the comparative strengths of humans and AI as well as human preferences. Such an approach ensures that AI's computational efficiency complements, rather than overrides, human capacity for creative and emotive expression. Additionally, collaboration design must consider differences in baseline performance, as writers with stronger baseline performance may experience greater adverse effects when excluded from creative responsibilities. To preserve high levels of creative diversity, organizations should retain significant human involvement or intentionally train and prompt AI systems differently to generate more diverse outputs. The preferred collaboration design will likely vary across organizational contexts. For example, organizations that depend on a broad range of ideas and perspectives may place greater weight on preserving diversity, whereas organizations performing more standardized work may prioritize efficiency. Educational settings may value learning and satisfaction, while media organizations may prioritize quality and speed. Our results suggest that collaboration design should be selected with these differing objectives in mind.

Moreover, our reflective feedback from participants indicates that AI can enhance human creativity in two distinct ways: as a collaborative partner in creative exploration and as a productivity enhancer that frees humans to focus on more creative subtasks. Specifically, 78.5% of participants emphasized idea generation and creativity support, while 58.5% highlighted AI's ability to streamline routine tasks like drafting, thereby freeing up time for more complex, creative thinking. Importantly, our results showed that humans should maintain significant creative roles for these enhancements to be achieved. One of our main recommendations for managers is to flexibly adapt the human–AI collaboration model to the baseline performance of the people using AI systems.

## Acknowledgments

We used Claude Opus 5.5 and Claude Fable 5.1 (Anthropic) and GPT-6 Sol and GPT-6 Astra (OpenAI) to assist with drafting, editing, and reviewing text, proofreading, and drafting reproducibility code. The authors reviewed and verified all content and take full responsibility for it.

## Supplementary Material

Supplementary material is available at *PNAS Nexus* online.

## Competing Interest

The authors declare that they have no competing interests.

## Funding

The authors declare no funding.

## Author Contributions

Conceptualization: K.H. and D.A.; Methodology: K.H. and D.A.; Investigation: D.A. and K.H.; Visualization: D.A.; Project administration: K.H. and D.A.; Supervision: K.H. and D.A.; Writing—original draft: K.H. and D.A.; Writing—review & editing: K.H. and D.A.

## Preprints

A preprint of this article is published at arXiv:2412.14199.

## Data Availability

All data and code used in the analysis are available at the Dataverse (https://doi.org/10.7910/DVN/ATXBCK).